\documentclass[prb,twocolumn,showpacs,amsmath,amssymb,%
floatfix,superscriptaddress,a4paper]{revtex4}

\def\ealla#1{{\rm e}^{#1}}
\def\--{\negthinspace - \negthinspace}
\usepackage{times} \normalfont 
\usepackage[T1]{fontenc} 
\usepackage{graphicx,color}
\usepackage{bm}

\begin{document}

\title{
  Rotational dynamics of CO solvated in small He clusters: 
  a quantum Monte Carlo study
}

\author{
  Paolo Cazzato
}
\email{
  cazzato@democritos.it
}
\affiliation{
  {\sl SMC} INFM -- Istituto Nazionale per la Fisica della Materia and
  Dipartimento di Fisica, Universit\`a di Roma {\sl La Sapienza} \\ 
  Piazzale Aldo Moro 2, I-00185 Rome, Italy
}

\author{
  Stefano Paolini
}
\email{
  paolini@sissa.it
}
\affiliation{
  SISSA -- Scuola Internazionale Superiore di Studi Avanzati and
  INFM-{\sl DEMOCRITOS} National Simulation Center, \\
  via Beirut 2--4, I-34014 Trieste,
  Italy
}
\author{
  Saverio Moroni
}
\email{
  moroni@caspur.it
}
\affiliation{
  {\sl SMC} INFM -- Istituto Nazionale per la Fisica della Materia and
  Dipartimento di Fisica, Universit\`a di Roma {\sl La Sapienza} \\ 
  Piazzale Aldo Moro 2, I-00185 Rome, Italy
}

\author{
  Stefano Baroni
}
\email{
  baroni@sissa.it
}
\affiliation{
  SISSA -- Scuola Internazionale Superiore di Studi Avanzati and
  INFM-{\sl DEMOCRITOS} National Simulation Center, \\
  via Beirut 2--4, I-34014 Trieste,
  Italy
}

\date{
  \today
}

\begin{abstract}
  The rotational dynamics of CO single molecules solvated in small He
  clusters (CO@He$_N$) has been studied using Reptation Quantum Monte
  Carlo for cluster sizes up to $N=30$. Our results are in good
  agreement with the roto-vibrational features of the infrared
  spectrum recently determined for this system, and provide a deep
  insight into the relation between the structure of the cluster and
  its dynamics. Simulations for large $N$ also provide a prediction of
  the effective moment of inertia of CO in the He nano-droplet regime,
  which has not been measured so far.
\end{abstract}


%% 36.40.-c Atomic and molecular clusters
%% 34.30.+h Intramolecular energy transfer; intramolecular dynamics; 
%%           dynamics of van der Waals molecules
%% 67.40.Yv Impurities and other defects
%% 36.40.Mr Spectroscopy and geometrical structure of clusters
%% 02.70.Ss Quantum Monte Carlo methods
%% 61.46.+w Nanoscale materials: clusters, nanoparticles, nanotubes, 
%%           and nanocrystal 
%% 33.20.Sn Rotational analysis
 
\maketitle

\section{Introduction}

Thanks to recent progresses in Helium nano-droplet isolation (HENDI)
spectroscopy,\cite{Scoles2001} the infrared and microwave spectra of
small molecules solvated in He$_N$ clusters are now becoming
accessible in the small- and intermediate-size
regimes.\cite{Jaeger2002,Jaeger2003,McKellar2003-OCS,McKellar2003-CO}
In the particular case of carbon monoxide, the roto-vibrational
spectrum of the molecule solvated in small He clusters (CO@He$_N$) has
been recently studied in the size range $N=2\--
20$.\cite{McKellar2003-CO} The infrared spectrum in the 2145 cm$^{-1}$
region of the C--O stretch consists of two $R(0)$ transitions which
smoothly correlate with the $a$-type ($K=0\leftarrow 0$) and $b$-type
($K=1\leftarrow 0$) $R(0)$ lines of the binary complex, CO@He$_1$. The
series of $b$-type transitions---which starts off about 7 times
stronger for $N=1$---progressively looses intensity as $N$ increases,
until it disappears around $N=7\-- 8$. Around this size, just before
it disappears, the $b$-type line seems to split in two. Analogously,
around $N=15$ the $a$-type line also seems to split, and the
assignment of experimental lines becomes uncertain for larger
clusters. Elucidating the relation existing between the position,
number, and intensity of the rotational lines and the size and
structure of the cluster is the principal goal of the present paper.

Computer simulations of quantum many-body systems have also
considerably progressed in recent years, allowing in some cases to
determine the low-lying spectrum of excited states. The rotational
dynamics of small molecules solvated in He clusters and nanodroplets
is one of these favorable instances. The scarcity of low-lying excited
states typical of superfluid systems makes it possible in this case to
extract information on the location and intensity of the spectral
lines from an analysis of the time series generated by quantum Monte
Carlo random walks.\cite{NoAntri-Cornell,NoAntri-PRL,POITSEH} The
rotational spectrum of OCS@He$_N$ has been studied along these lines
in Refs. [\onlinecite{Whaley2003,NoAntri-2003}]. Among the many
different flavors of quantum Monte Carlo available in the literature,
we adopt {\sl Reptation Quantum Monte
  Carlo}\cite{NoAntri-Cornell,NoAntri-PRL} which we believe presents
distinctive advantages in the present case and which will be briefly
introduced in Sec. \ref{sec:RQMC}. In Sec. \ref{sec:results} we
present and discuss our results, whereas Sec. \ref{sec:conclusions}
contains our conclusions.

\section{Theory, algorithms, and technical details} \label{sec:RQMC}

Virtually all the ground-state quantum simulation methods are based on
the prior knowledge of some approximate wave-function, $\Phi_0$, for
the system under study. In the Variational Monte Carlo Method (VMC)
one contents oneself of this knowledge and the simulation simply aims
at calculating the complicated multi-dimensional integrals which are
needed to estimate ground-state {\sl approximate} expectation values,
$ \langle \Phi_0 | \widehat A | \Phi_0 \rangle $ (here and in the
following quantum-mechanical operators are indicated with a hat,
$\widehat{~~}$). To this end, a random walk in configuration space is
generated according to the Langevin equation:
\begin{equation}
  d{\textbf x} = \epsilon ~ {\textbf f}_0({\textbf x}) + d\bm{\xi},
  \label{eq:langevin}
\end{equation}
where ${\textbf x} \equiv \{x_\alpha\} $ indicates the coordinates of the
system, $\epsilon$ is the step of time discretization,
\begin{equation}
  {\textbf f}_0({\textbf x}) = 2 {\partial \log\left ( \Phi_0({\textbf
        x}) \right ) \over  \partial {\textbf x} },  
\end{equation}
and $d\bm{\xi}$ is a Gaussian random variable of variance $2\epsilon$:
$\langle d\xi_\alpha d\xi_\beta \rangle = 2\epsilon \delta_{\alpha
 \beta}$. {\sl Approximate} ground-state quantum expectation values 
are then estimated
as time averages over the random walk, Eq. (\ref{eq:langevin}).

Within Reptation Quantum Monte Carlo (RQMC), {\sl exact} ground-state
expectation values and imaginary-time correlation functions are
calculated as appropriate derivatives of the pseudo partition
function, in the low temperature (large $T$) limit:
\begin{equation}
  {\cal Z}_0 = \langle\Phi_0|e^{-T \widehat H}|\Phi_0\rangle ,
  \label{eq:Z0}
\end{equation}
where $\widehat H$ is the Hamiltonian of the system. By breaking the
time $T$ into $P$ intervals of length $\epsilon = T/P$, Eq.
(\ref{eq:Z0}) can be given a path-integral representation:
\begin{equation}
  {\cal Z}_0 \approx \int \Phi_0({\textbf x}_0)
  \Pi_{i=0}^{P-1} \langle {\textbf x}_i|
  e^{-\epsilon\widehat H} |{\textbf x}_{i+1} \rangle
  \Phi_0({\textbf x}_P) d^{P+1}{\textbf x}~. \label{eq:PI}
\end{equation}
For the relatively small systems considered here, it is sufficient to
use the {\em primitive approximation} to the imaginary-time
propagator: 
\begin{equation}
  \langle {\textbf x}|e^{-\epsilon\widehat H}|{\textbf y}
  \rangle \propto 
  \ealla{
    -({\textbf x}-{\textbf y})^2/2\epsilon
    -\epsilon[V({\textbf x})+V({\textbf y})]/2
  }
  +{\cal O}(\epsilon^3),
\end{equation}
where $V({\textbf x})$ is the potential energy at point ${\textbf x}$.
The dynamical variables of the statistical-mechanical system whose
partition function is given by Eq. (\ref{eq:PI}) are segments of the
VMC random walk generated from Eq. (\ref{eq:langevin}), ${\textbf
  x}(\tau)$ of length $T$, which we call {\sl reptiles}. As the random
walk proceeds, the reptile is allowed to creep back and forth: new
configurations of the reptile are accepted or rejected according to a
Metropolis test made on the integrand of the path-integral
representation of ${\cal Z}_0$, Eq. (\ref{eq:PI}). It can be
demonstrated\cite{NoAntri-Cornell,NoAntri-PRL} that in the large-$T$
limit the sample of reptile configurations thus generated is such that
the sample average of quantities like
\begin{equation}
  {\cal A}[{\textbf x}(\tau)] = {1\over T} \int_0^T A({\textbf
  x}(\tau)) d\tau 
  \label{eq:A_estimator}
\end{equation}
converges without any systematic bias (but those due to the finite
values of $\epsilon$ and $T$) to $\langle \widehat A \rangle = \langle
\Psi_0 | \widehat A | \Psi_0 \rangle$, $\Psi_0$ being the {\sl exact}
ground-state wavefunction of the system. Even more interesting is the
fact that sample averages of reptile time correlations,
\begin{equation}
  {\cal C}_A(\tau) = {1\over T-\tau} \int_0^{T-\tau} A({\textbf x}(\tau')) 
  A({\textbf x}(\tau'+\tau)) d\tau', \label{eq:CA_estimator}
\end{equation}
provide equally unbiased estimates of the corresponding quantum
correlation functions in imaginary time: 
\begin{eqnarray}
  C_{\widehat A}(it) &= & 
  \langle \Psi_0 | \widehat A(it)
  \widehat A(0) | \Psi_0 \rangle \nonumber \\
  &\equiv&  
  \langle \Psi_0 | \ealla{\widehat H t} \widehat A
  \ealla{-\widehat H t} \widehat A | \Psi_0 \rangle,
\end{eqnarray}
${\cal C}_A(\tau)\approx C_{\widehat A}(i\tau)$. The absorption
spectrum of a molecule solvated in a non polar environment is given by
the Fourier transform of the autocorrelation function of its electric
dipole, $\textbf d$:
\begin{eqnarray}
  I(\omega) &\propto& 2\pi \sum_n | \langle \Psi_0|\widehat {\textbf
    d}|\Psi_n \rangle |^2 \delta(E_n-E_0-\omega) \nonumber \\ &=&
  \int_{-\infty}^\infty \ealla{i\omega t} \langle \widehat {\textbf d}(t)
  \cdot \widehat {\textbf d} (0) \rangle dt,
\end{eqnarray}
where $\Psi_0$ and $\Psi_n$ are ground- and excited-state
wavefunctions of the system respectively, and $E_0$ and $E_n$ the
corresponding energies. The dipole of a linear molecule---such as
CO---is oriented along its axis, so that the optical activity is
essentially determined by the autocorrelation function of the
molecular orientation versor: $c(t)\equiv C_{\widehat {\textbf n}}
(t)=\langle \Psi_0| \ealla{i\widehat Ht} \widehat {\textbf n}
\ealla{-i\widehat Ht} \widehat {\textbf n} |\Psi_0 \rangle$. We have
seen that RQMC gives easy access to the analytic continuation to
imaginary time of correlation functions of this kind. From now on,
when referring to {\sl time correlation functions}, we will mean {\sl
  reptile time correlations}, {\sl i.e.} quantum correlation functions
in imaginary time. Continuation to imaginary time transforms the
oscillatory behavior of the real-time correlation function---which is
responsible for the $\delta$-like peaks in its Fourier
transform---into a sum of decaying exponentials whose decay constants
are the excitation energies, $E_n-E_0$, and whose spectral weights are
proportional to the absorption oscillator strengths, $|\langle
\Psi_0|{\textbf d}|\Psi_n \rangle |^2$. Dipole selection rules imply
that only states with $J=1$ can be optically excited from the ground
state which has $J=0$. Information on excited states with different
angular momenta, $J$, can be easily obtained from the multipole
correlation functions, $c_J(\tau)$, defined as the reptile time
correlations of the Legendre polynomials:
\begin{eqnarray}
  c_J(\tau) 
  &=& 
  \left \langle 
    P_J ( {\textbf n}(\tau) \cdot {\textbf n}(0) )
  \right \rangle 
  \nonumber \\
  &\equiv& 
  \left \langle 
    {4\pi\over 2J+1} \sum_{M=-J}^J Y^*_{JM}\bigl
    ({\textbf n}(\tau) \bigr ) Y_{JM}\bigl ({\textbf n}(0) ) 
  \right \rangle 
  \quad {.} \quad 
\end{eqnarray}

Both the He-He and the He-CO interactions used here are derived from
accurate quantum-chemical calculations.\cite{Korona,Heijmen} The CO
molecule is allowed to perform translational and rotational motions,
but it is assumed to be rigid. The trial wavefunction is chosen to be
of the Jastrow form:
\begin{equation}
  \Phi_0 = \exp \left [ -\sum\limits_{i=1}^N {\cal U}_1(r_i,\theta_i)
    -\sum\limits_{i<j}^N {\cal U}_2(r_{ij}) \right ], 
\end{equation}
where ${\textbf r}_i$ is the position of the i-th atom with respect to
the center of mass of the molecule, $r_i=|{\textbf r}_i|$, $\theta_i$
is the angle between the molecular axis and ${\textbf r}_i$, and
$r_{ij}$ is the distance between the i-th and the j-th helium atoms.
${\cal U}_1$ is expressed as a sum of five products of radial
functions times Legendre polynomials. All radial functions (including
${\cal U}_2$) are optimized independently for each cluster size with
respect to a total of 27 variational parameters. The propagation time
is set to $T=\rm 1~ K^{-1}$, with a time step of $\epsilon = \rm
10^{-3}~ K^{-1}$. The effects of the length of the time step and of
the projection time have been estimated by test simulations performed
by halving the former or doubling the latter. These effects were
barely detectable on the total energy, and very small on the
excitation energies discussed below (we estimate that more converged
simulations would actually improve the already excellent agreement
with experimentally observed spectra).

\begin{figure}[b]
  \hbox to \hsize{\hfill
    \includegraphics[width=75mm]{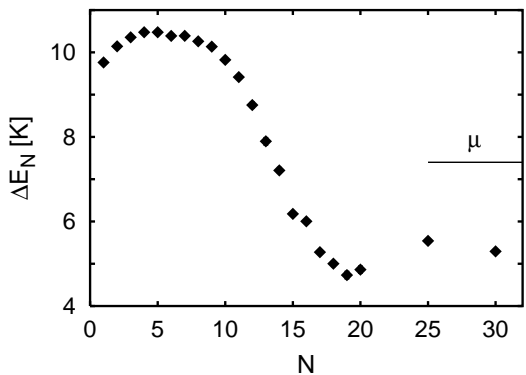}
    \hfill
  }
  \caption{
    Atomic binding energy, $\Delta E_N=E_N-E_{N-1}$, as a function of
    the cluster size in CO@He$_N$. The horizontal line on the right of
    the figure indicates the chemical potential in bulk $^4$He,
    $\mu=7.4~ \rm K$ [\onlinecite{NoAntri-PRL}]. 
    \label{fig:deltaE} }
\end{figure}

The estimate of excitation energies and spectral weights from
imaginary-time correlations amounts to performing an inverse Laplace
transform, a notoriously ill-conditioned operation which is severely
hindered by statistical noise.\cite{gubernatis} For each value of $J$,
we extract the value of the two lowest-lying excitation energies,
$\epsilon_{a,b}^J$ ---{\sl i.e.} the two smallest decay constants in
$c_J(\tau)$---as well as the corresponding spectral weights,
$A_{a,b}^J$, from a fit of $c_J(\tau)$ to a linear combination of
three decaying exponentials. This fitting procedure does not solve in
general the problem of obtaining the spectrum from a noisy
imaginary-time correlation function. However, if we know in advance
that very few strong peaks, well separated in energy, nearly exhaust
the entire spectral weight, their position and strength can be
reliably estimated from this multi-exponential fit. In the present
study these favorable conditions are usually met, although the
limitations of the procedure will show in some cases, as discussed
below.

\begin{figure}
  \begin{center}
    \hspace{-15mm}
    \includegraphics[width=100mm]{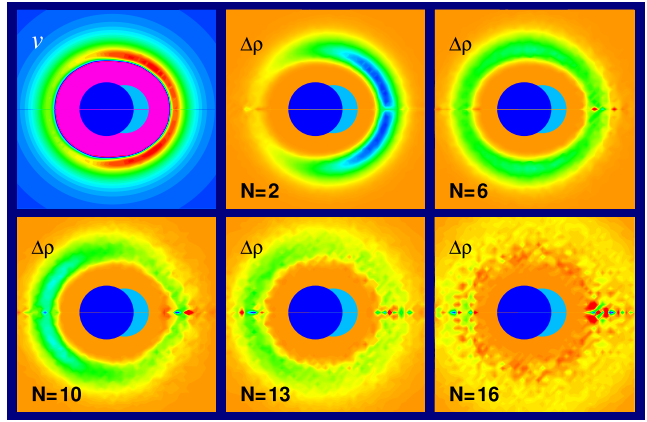}
  \end{center}
  \caption{
    (color) Upper left panel: He-CO interaction potential. C (blue)
    and O (cyan) atoms are pictured by two circles whose radius is the
    corresponding Van de Waals radius. The other panels picture the
    differential He density, $\Delta\rho_N = \rho_N-\rho_{N-1}$ for
    various sizes of the CO@He$_N$ cluster. Color convention is
    rainbow: red to purple in order of increasing magnitude.
    \label{fig:effetti-speciali}
  }
\end{figure}

\section{Results and discussion} \label{sec:results}

RQMC simulations have been performed for CO@He$_N$ clusters in the
size range $N=1\-- 30$. In Fig. \ref{fig:deltaE} we report the values
of the He atomic binding energy, $\Delta E_N=E_{N-1}-E_N$, as a
function of the cluster size. $\Delta E_N$ first increases up to
$N=4\-- 5$, and it stays roughly constant in the range $N=5\-- 8$;
from this size on $\Delta E_N$ starts decreasing, first slowly, then,
from $N=10\-- 11$, rapidly down to a minimum at $N=19$. For $N>19$
$\Delta E_N$ increases again and slowly tends to the nanodroplet
regime (where it coincides with the bulk chemical potential, $\mu=7.4~
\rm K$~~\cite{NoAntri-PRL}) which is however attained for much larger
cluster sizes than explored here.\cite{paesani} This behavior can be
understood by comparing the shape of the CO--He potential energy
function, $v({\textbf r})$, with the incremental atomic density
distributions, $\Delta\rho_N({\textbf r}) = \rho_N({\textbf r}) -
\rho_{N-1}({\textbf r})$, where $\rho_N$ is the expectation value of
the He density operator:
\begin{equation}
  \widehat \rho({\textbf r}) = \sum_{i=1}^N \delta({\textbf
    r}-{\textbf r}_i)
\end{equation}
(see Fig. \ref{fig:effetti-speciali}). For very small $N$ the atomic
binding energy is dominated by the He--CO attraction which is
strongest in a well located atop the oxygen atom. As He atoms fill
this well, $\Delta E_N$ first slightly increases, as a consequence of
the attractive He--He interaction, then, for larger $N$, the increased
He--He interaction is counter-balanced by the spill-out of He atoms
off the main attractive well, until for $N\approx 9$ the reduction of
the He--CO interaction overcomes the increased attraction and the
binding energy starts decreasing steeply. For $N$ in the range 10--14
He density accumulates towards the C pole, while, around $N=15$, the
first solvation shell is completed and the differential atomic
density, $\Delta\rho_N$, is considerably more diffuse starting from
$N=16$. $\Delta E_N$ reaches a minimum at $N=19$. For larger sizes,
the trend in the atomic binding energy is dominated by the increase of
the He--He attraction related to the increase of the cluster size,
until it will converge to the bulk chemical potential.

\begin{figure}[t]
  \hbox to \hsize{\hfill
    \includegraphics[width=75mm]{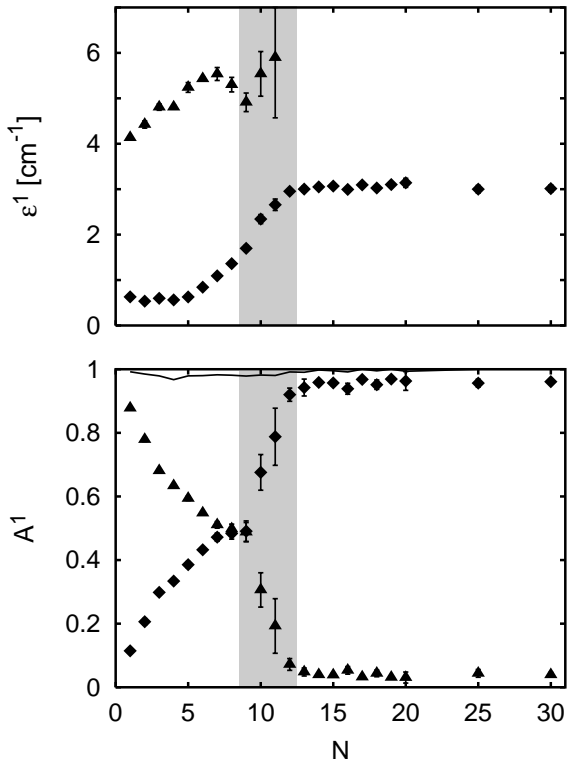}
    \hfill
  }
  \caption{
    Upper panel: position of the rotational lines of CO@He$_N$ as
    obtained from RQMC simulations as a function of the cluster size,
    $N$. $a$-type lines are indicated with triangles, $b$-type lines
    with diamonds. Lower panel: spectral weights of the lines reported
    in the lower panel; the continuous line near the upper border of
    the figure corresponds to the sum of the spectral weights.  }
  \label{fig:EA_QMC}
\end{figure}

\begin{figure}[t]
  \hbox to \hsize{\hfill
    \includegraphics[width=75mm]{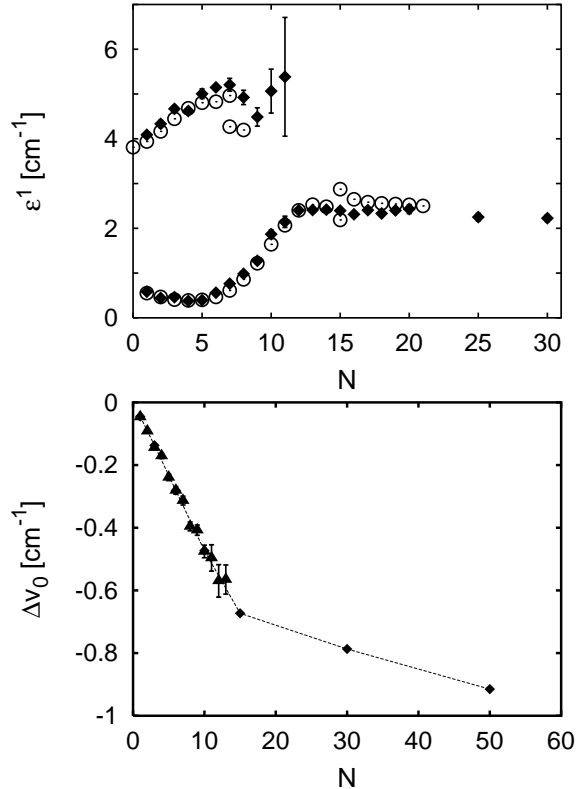}
    \hfill
  }
  \caption{
    Upper panel: positions of the infrared lines of CO@He$_N$ as
    observed experimentally (Ref. \onlinecite{McKellar2003-CO}, empty
    circles) and as estimated from the present simulations and
    corrected by the estimated vibrational shift (solid diamonds, see
    text). Lower panel: vibrational shift of the lines, as estimated
    in the present work (triangles) and in Ref. \onlinecite{paesani}
    (diamonds and dashed line). 
  }
  \label{fig:E_expt}
\end{figure}
 
In Fig. \ref{fig:EA_QMC} we report the positions and spectral weights
of the rotational lines, as functions of the cluster size, $N$. In the
size range $N=1\-- 9$, analysis of the dipole time correlations
clearly reveals the presence of two peaks, with the weight of the
higher--energy ($b$-type) rapidly decreasing by almost a factor 2.
Note that the sum of the spectral weights of these two lines nicely
sums to one, indicating that they exhaust all the oscillator strength
available for optical transitions originating from the ground state.
For $N$ between 10 and 12 (shaded area in Fig. \ref{fig:EA_QMC}) the
situation is less clear. As the weight of the $b$-type line drops to
zero, the statistical noise on its position grows enormously.
Furthermore the multi-exponential fit introduces some ambiguity, as
the results are somewhat sensitive to the number of terms in the sum.
However the important information that one line disappears between
$N=10$ and 12 is clear. For larger $N$ only one relevant line remains,
and the robustness of the fitting procedure is recovered, with the
minor exception of the sizes around $N=16$, where the minimum of the
$\chi^2$ appears to be less sharp, possibly correlating with the
splitting of the line observed in the infrared spectra for $N=15$ (see
below).

In the upper panel of Fig. \ref{fig:E_expt} we compare the rotational
structure of the observed infrared (vibrational)
spectrum\cite{McKellar2003-CO} with the rotational excitation energies
calculated in this work. Experimental data are referred to the center,
$\nu_0$, of the vibrational band for $N=0$ (CO monomer). In order to
better compare our predictions with experiments, we have corrected the
former with an estimate of the {\sl vibrational shift}, $\Delta
\nu_0$, i.e. the displacement of the vibrational band origin as a
function of the number of He atoms. The vibrational shift can be
calculated as the difference in the total energy of the cluster
obtained with two slightly different potentials,\cite{Heijmen}
$v_{00}$ and $v_{11}$, representing the interaction of a He atom with
the CO molecule in its vibrational ground state and first excited
states, respectively. Our estimate of the vibrational shift as a
function of the cluster size is reported in the lower panel of Fig.
\ref{fig:E_expt}. Since the evaluation of a small difference between
two large energies is computationally demanding for large clusters,
$\Delta \nu_0$ has been evaluated perturbatively with respect to the
difference $v_{00}-v_{11}.$\cite{paesani} We have used the vibrational
shift calculated in Ref.~\onlinecite{paesani} after verifying on small
clusters that the perturbative treatment is reliable. The agreement
between our results and experiments is remarkable. Some of the
features of the observed spectrum, however, call for a deeper
understanding and theoretical investigation. Two questions, in
particular, naturally arise. Why two peaks are observed in the
small-size regime, and what determines the disappearance of one of
them at $N=8$? What determines the split of the higher-frequency
($b$-type) line at $N=7$ and of the lower-frequency ($a$-type) one at
$N=15$?

\begin{figure}
  \hbox to \hsize{\hfill
    \includegraphics[width=75mm]{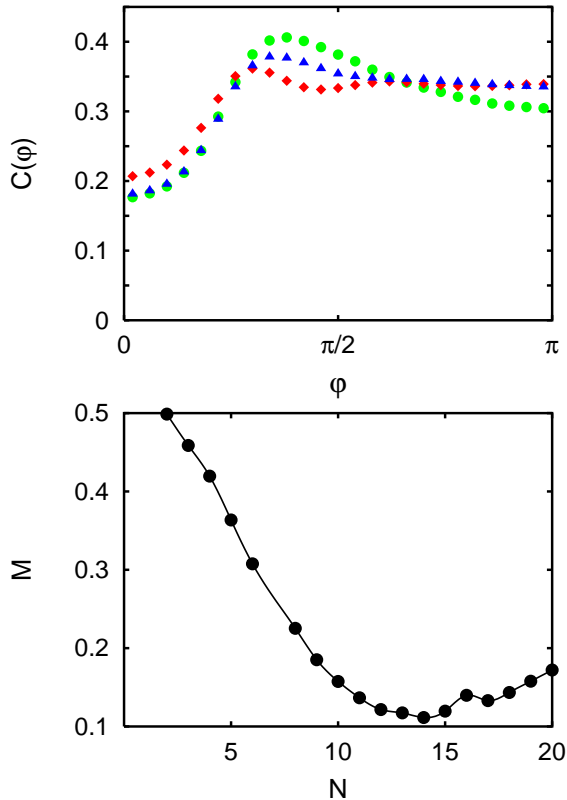}
    \hfill
  }
  \caption{
    \label{fig:montarozzi} 
    (color) Upper panel: probability density of finding two He atoms
    which form a dihedral angle $\phi$ with respect to the molecular
    axis; the probability is normalized to 1.  Results pertain to
    clusters with $N=3$ (green circles), $N=6$ (blue triangles) and
    $N=13$ (red diamonds). Lower panel: the integrated probability
    density, defined in Eq.~(\ref{eq_m}), as a function of the cluster
    size.  }
\end{figure}

The existence of two lines for small $N$ is likely due to a larger
asymmetry of the cluster in this regime. If the CO@He$_N$ complex is
described as a rigid rotor, in fact, one would have one rotational
line originating from a $J=0$ ground state if the complex has
cylindrical symmetry, while this line would double if some of the
atomic density accumulates in a longitudinal protrusion. The inertia
of the complex would in this case be larger for a rotation about an
axis perpendicular to a plane containing the protrusion ({\sl
  end-over-end rotation}) than about an axis lying on such a plane.
Given that the He density in the ground state of CO@He$_N$ is
cylindrically symmetric, any departure from this symmetry can only
show up in higher correlation functions. The situation is conceptually
similar to that of a fluid whose density is homogeneous and whose
structure at the atomic scale is reflected in the pair correlation
function. Analogously, we define an atomic angular correlation
function, $C(\phi)$, as the probability of finding two He atoms which
form a dihedral angle $\phi$ with respect to the molecular axis:
\begin{equation}
  C(\phi) = 
  \left \langle
    {1\over N(N-1)} \sum_{i\ne j} \delta(\phi_i-\phi_j-\phi) 
  \right \rangle.
\end{equation}
In the upper panel of Fig. \ref{fig:montarozzi} we show $C(\phi)$, for
different cluster sizes. The depletion of $C$ for $\phi$ larger than
$\pi/2$, clearly visible for $N=3$ (green circles), indicates a
tendency of the He atoms to cluster on a same side of the molecular
axis. For larger clusters, however, this effect weakens to the extent
that it becomes difficult to disentangle from the structural
information related to the He--He interaction (the dimple at small
$\phi$ and the subsequent maximum around $\phi=0.3\-- 0.4~\pi$). A
more sensitive measure of the propensity of He atoms to cluster on a
side of the molecule is given by the integral of $C(\phi)$ from 0 to
$\frac{\pi}{2}$,
\begin{equation}
  M = \int_0^{\pi\over 2} C(\phi) d\phi-{1\over 2}.
\label{eq_m}
\end{equation}
In the lower panel of Fig. \ref{fig:montarozzi} we display $M$ as a
function of the cluster size, $N$: one sees that $M$ decreases with
$N$ and reaches a minimum at $N=14$. This is the size at which the
first solvation shell is completed, and the cluster asymmetry
increases again when the second shell starts to build. The rotational
spectrum of the solvated molecule, however, is insensitive to this
asymmetry for clusters of this and larger sizes because the motion of
He atoms in the second and outer solvation shells is decoupled from
that in the first and from molecular rotation. The existence of a
longitudinal asymmetry is a necessary condition for the doubling of
the rotational line. Whether or not this condition is also sufficient
depends on the dynamics: if quantum fluctuations make the motion of
the protrusion around the molecular axis fast with respect to the
molecular rotation, then the asymmetry is effectively washed out. The
existence of two lines in the rotational spectrum of the molecule
implies therefore that an asymmetry in the {\em classical}
distribution of He atoms around the molecular axis exists; that the
molecular inertia is sensitive to this asymmetry (the protrusion can
be `dragged' along the molecular rotation); and that the motion of
this protrusion around the molecular axis is not adiabatically
decoupled from the molecular rotation.

\begin{figure}
  \hbox to \hsize{
    \hfill
    \includegraphics[width=75mm]{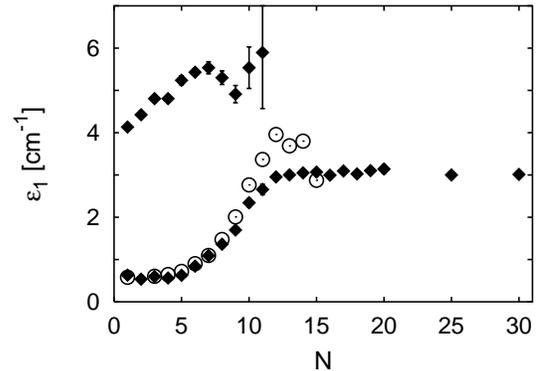}
    \hfill
  }
  \caption{
    Diamonds: CO rotational frequencies in CO@He$_N$ as
    functions of the cluster size, $N$ (same as in
    Fig. \ref{fig:EA_QMC}). Dots: frequency of the lowest mode
    appearing in the spectral analysis of the angular He--He
    correlation function (see Eq. (\ref{eq:cuu})).
  }
  \label{fig:adiab}
\end{figure}

In order to better characterize the motion of He atoms around the
molecule and the coupling of this motion to molecular rotation, we
examine the imaginary-time correlations of the versor, $\mathbf{u}$,
of the He center of mass, $\mathbf{r}_{CM}$, relative to the molecular
center of mass:
\begin{equation}
  {\cal C}_{\mathbf{u}}(\tau)=
  \langle 
  \mathbf{u}(\tau)\cdot \mathbf{u}(0) 
  \rangle. \label{eq:cuu}
\end{equation}
For the binary complex, He--CO, $\mathbf{r}_{CM}$ coincides with the
position of the helium atom, and we expect its angular dynamics to be
strongly correlated to the molecular rotation, at least in the
end-over-end mode. In Fig. \ref{fig:adiab} we report the frequency of
the slowest mode appearing in the spectral analysis of ${\cal
  C}_{\mathbf{u}}(\tau)$, $\epsilon_{\mathbf{u}}$, as a function of
$N$, and compare it with the corresponding frequencies of the
molecular rotation. We see that for cluster sizes up to $N=9\-- 10$,
$\epsilon_{\mathbf{u}}$ is degenerate with the $a$-type frequency in
the molecular rotational spectrum, with a spectral weight which passes
from $A_{\mathbf{u}} \approx 1$ for $N=1$ to $A_{\mathbf{u}} \approx
0.7$ for $N=10$. These findings are a manifestation of the fact that
He atoms are dragged along the slowest, end-over-end, rotation of the
solvated molecule, and that the effect of this dragging decreases when
more He states with $J=1$ become available and subtract spectral
weight to the slowest mode. For $N>10$, $\epsilon_{\mathbf{u}}$
further increases and departs from $\epsilon_a$, indicating an
effective decoupling of the two kinds of motion. In this regime, the
effective rotational constant $B$ of the solvated molecule is almost
independent of the cluster size. Free molecular rotation with an
increased moment of inertia with respect to the gas phase is the
typical signature of superfluid behavior in He nanodroplets.
Extrapolating the result obtained for $N$ up to 30 to the nanodroplet
limit, we predict a renormalization factor of the $B$ value of 0.78.
The lowest atomic mode, $\epsilon_{\mathbf{u}}$, slows down again for
$N=15$. This is due, however to the slow He motion in the second
solvation shell which hardly affects the rotation of the solvated
molecule. Although the resolution that can be achieved with our
simulations is not sufficient to detect the doubling of the $a$ and
$b$ lines which is experimentally observed for $N=15$ and $N=7$
respectively, it is interesting to notice that the former occurs in
correspondence with the crossing between $\epsilon_{\mathbf{u}}$ and
$\epsilon_a$, possibly due to the resonant interaction between the two
modes. It is tempting to assume that a similar mechanism may be
responsible for the doubling of the $b$ line at $N=7$, involving
however higher-energy He states. A deeper study of the He dynamics
would clarify this point.

\section{Conclusions} \label{sec:conclusions}
Computer simulations of quantum many-body systems have reached such a
degree of sophistication and reliability that in some cases they can
be used to provide information, complementary to that which can be
obtained in the laboratory, on the dynamical processes probed
spectroscopically.

In the case of small polar molecules solvated in He clusters, for
instance, the calculation of the time autocorrelation of the molecular
dipole (which is the quantity directly coupled to the experimental
probe) allows to reproduce rather accurately the roto-vibrational
excitation energies which are now becoming experimentally accessible
for small clusters ($N=1\--20$). Even more importantly, computer
simulations give direct access to quantitities and features (such as,
{\sl e.g.}, static and dynamic properties of the He matrix) which are
not accessible to the experiment, and whose knowledge provides the
basis for understanding the relation between structure and dynamics in
these confined boson systems.

In the specific case of CO@He$_N$, which is the subject of the present
study and of a recent infrared spectroscopy
experiment,\cite{McKellar2003-CO} the presence of two spectral
lines---$a$-type and $b$-type, evolving respectively from the
end-over-end and from the free-molecule rotations of the binary
complex---is related to the propensity of the He atoms to cluster on a
same side of the molecular axis, which we measure by an angular pair
distribution function: as more He atoms progressively fill the first
solvation shell, their clustering propensity weakens; the CO impurity
gets more isotropically coated, looses a preferred axis for the
free-molecule mode, and the $b$--type line disappears.

The time autocorrelation of the versor of the He center of mass
provides dynamical information on the He atoms in excited states with
$J=1$. We find a substantial spectral weight on a He mode whose
energy, $\epsilon_{\mathbf{u}}$, is degenerate with the $a$-type line
for $N$ up to about ten. This indicates that some of the He density is
dragged along by the molecular rotation---in other words, part of the
angular momentum in the cluster mode involving molecular rotation is
carried by the He atoms. We also find that for larger clusters the
molecular rotation effectively decouples from this He mode, and its
energy $\epsilon_a$ becomes essentially independent of the number of
He atoms. Based on the nearly constant value of $\epsilon_a$ in the
range of $N$ between 15 and 30, well beyond completion of the first
solvation shell, we predict the effective rotational constant in the
nanodroplet limit to be smaller by a factor 0.78 than its gas phase
value. 

\vbox to 0pt{\vss}

\begin{acknowledgments}
  We would like to thank A.R.W. McKellar for providing us with a
  preprint of Ref. [\onlinecite{McKellar2003-CO}] prior to
  publication. We are grateful to G. Scoles for bringing that work to
  our attention, for his continuous interest in our work, and for many
  useful discussions. Last but not least, we wish to thank S. Fantoni
  for his encouragement and for a critical reading of our manuscript.
  This work has been partial supported by the Italian {\em MIUR}
  through {\em PRIN}.
\end{acknowledgments}

\end{document}